\documentclass[journal]{IEEEtran}
\usepackage[T1]{fontenc}

\ifCLASSINFOpdf
\else
\fi
\usepackage{amsmath}
\usepackage{amsthm}
\usepackage{amsfonts}
\usepackage{float}
\usepackage{pgfplots}
\usepackage{tikz}
\usepackage[colorinlistoftodos]{todonotes}
\usetikzlibrary[pgfplots.groupplots]

\newlength{\hsep}

\usepackage{fancyhdr}
\usepackage{lipsum}
\usepackage{fancyhdr}
\fancypagestyle{title}{%
  \setlength{\headheight}{22pt}%
  \fancyhf{}
  \lhead{PREPRINT, MARCH 14, 2022}
\rhead{\thepage}
}%

\pgfplotsset{width=8.5cm,compat=1.17}

\newtheorem{theorem}{Theorem}[section]
\newtheorem{property}[theorem]{Property}

\begin{document}

\title{Pulses with Minimum Residual Intersymbol Interference for Faster than Nyquist Signaling}

\author{Youssef~Jaffal,  Alex~Alvarado, \textit{Senior Member, IEEE}

\thanks{This work has  received  funding  from  the  European  Research  Council (ERC) under the European Union’s Horizon 2020 research and innovation programme under Grant 757791. \textit{(Corresponding author: Youssef Jaffal.)}

The authors are with the Department of Electrical
Engineering, Eindhoven University of Technology, 5600 MB Eindhoven,
The Netherlands (e-mail: y.jaffal@tue.nl; a.alvarado@tue.nl).

}
}

\maketitle
\thispagestyle{title}
\begin{abstract}
Faster than Nyquist signaling increases the spectral efficiency of pulse amplitude modulation by accepting inter-symbol interference, where an equalizer is needed at the receiver. Since the complexity of an optimal equalizer increases exponentially with the number of the interfering symbols, practical truncated equalizers assume shorter memory. The power of the resulting residual interference depends on the transmit filter and limits the performance of truncated equalizers. In this paper, we use numerical optimizations and the prolate spheroidal wave functions to find optimal time-limited pulses that achieve minimum residual interference. Compared to root raised cosine pulses, the new pulses decrease the residual interference by an order of magnitude, for example, a decrease by 32 dB is achieved for an equalizer that considers four interfering symbols at $57\%$ faster transmissions. As a proof of concept, for the $57\%$ faster transmissions of binary symbols, we showed that using the new pulse with a $4$-state equalizer has better bit error rate performance compared to using a root raised cosine pulse with a $128$-state equalizer.
\end{abstract}


%
\IEEEpeerreviewmaketitle

\section{Introduction}
\label{Sect1}

Faster than Nyquist (FTN) was introduced by Mazo in 1975 \cite{mazo1975faster}, where he found that increasing the signaling rate of sinc pulses by $25\%$ does not affect the minimum distance between binary coded codewords, which allows to increase the
data rates without affecting the probability of error when using an optimum receiver. Almost 30 years later, the study of minimum distance in FTN was extended to root raised cosine (RRC) filters in \cite{liveris2003exploiting}. Constrained theoretical information rates of FTN were investigated in \cite{rusek2009constrained} and \cite{jaffal2019achievable}, where it was shown that FTN is beneficial from an information theoretic perspective. 

As FTN signaling causes intersymbol interference (ISI), achieving the full gains that are suggested by the analysis of the minimum distance and the theoretical rates may require high-complexity (or  impractical) receivers. Optimal receivers with reduced complexity were proposed in \cite{forney1972maximum} and \cite{ungerboeck1974adaptive} for channels with ISI. Forney proposed to use a whitening filter followed by a Viterbi algorithm in \cite{forney1972maximum}. In \cite{ungerboeck1974adaptive}, Ungerboeck proposed a modified Viterbi algorithm that operates directly on the received symbols without using a whitening filter. 

As the complexity of the optimum equalizer increases exponentially with the number of interfering symbols, the equalizer needs to be truncated in order to be used in practical systems \cite{mclane1980residual}. 
In \cite{liveris2003exploiting}, practical ways of coding and equalization were proposed to achieve some gains of FTN, where a truncated modified Viterbi algorithm (TMVA) was proposed to detect the transmitted symbols. Truncated equalizers offer a trade-off between complexity and probability of error, where the effect of the residual ISI (RISI) can be decreased at the cost of increased complexity. The main issue of RISI is the error floor at high signal to noise ratio, since the signal to RISI ratio remains constant \cite{liveris2017fundamentals}.  Some approaches to mitigate the effect of RISI were proposed in the literature such as the decision feedback cancellation \cite{liveris2017fundamentals}. However, such methods increase the complexity of the system. The RISI depends on the used pulse shaping filter, for example, \cite[Table 1]{liveris2017fundamentals} shows the ISI coefficients that are obtained from the RRC pulses.

In this paper, we take a different approach and we propose to mitigate RISI by using pulse shaping filters that minimize the RISI power. Our proposal does not increase the complexity of FTN systems as it requires only replacing the  RRC pulse by the pulses that are tailored for FTN regimes. We find optimal time-limited pulses that minimize the RISI power while having a duration and an out-of-band energy (OOBE) that are equal to the ones of the truncated RRC. We propose a numerical method that makes use of the prolate spheroidal wave functions (PSWFs). The PSWFs form a complete orthonormal set for time-limited functions and they were used in \cite{halpern1979optimum} and \cite{jaffal2019achievable} to design optimal time-limited pulses. The main contribution of this paper is to introduce new pulses that have the ability to improve the performance  and the complexity of the receiver in FTN systems.

The paper is organized as follows: in Sec. \ref{Sec:SystemModel} we present the considered system model. In Sec. \ref{Sec:ProbForm} we formulate our optimization problem and we propose a numerical solution to search for the optimal pulses. In Sec. \ref{Sec:NumOpt} we present the results of the numerical optimizations and simulations. And we finally conclude in Sec. \ref{Sec:Conc}.

\section{System Model}
\label{Sec:SystemModel}

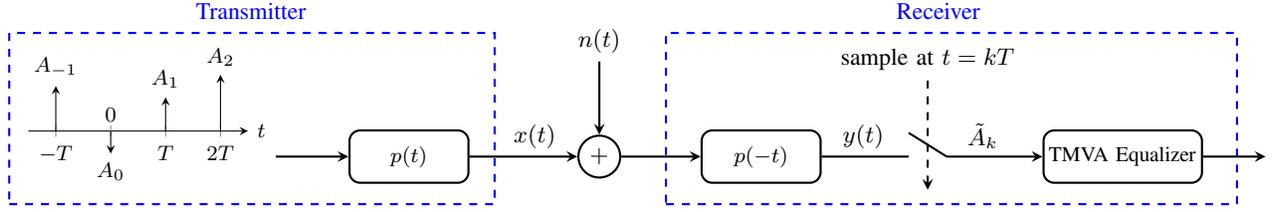
\begin{figure*}[t!]
	\centering
		\setlength{\hsep}{8ex} 
\begin{tikzpicture}[
font=\small,
>=stealth,auto,
block/.style={rectangle,thick,draw,inner sep=2pt,minimum width=1.25\hsep, minimum height=0.55\hsep,font=\footnotesize,rounded corners},
line1/.style = {draw, thick,->,rounded corners}
]
\coordinate (C);
\coordinate[left=2\hsep of C] (L1);
\coordinate[left=1.4\hsep of L1] (L2);
\coordinate[right=1.25\hsep of C] (R1);
\coordinate[right=1.7\hsep of C] (R2);
\coordinate[right=3.25\hsep of C](Switch_L);
\coordinate[above=0.25\hsep of Switch_L](Switch_U);
\coordinate[right=3.65\hsep of C](Switch_R);
\coordinate[right=5.5\hsep of C](Equalizer);
\coordinate[above=1\hsep of C] (U1);
\coordinate[right=7\hsep of C] (Rend);
\coordinate[right=3.45\hsep of C](dash_UU);
\coordinate[above=0.8\hsep of dash_UU](dash_U);
\coordinate[below=0.35\hsep of dash_UU](dash_B);

\draw[blue,thick,dashed] (-6.2\hsep,-0.5\hsep) rectangle (-1.1\hsep,1.3\hsep);
\draw[blue,thick,dashed]  (0.7\hsep,-0.5\hsep) rectangle (6.7\hsep,1.3\hsep);

\node[circle,inner sep=0.5ex,right=0cm of C, draw, thick,fill=white,anchor=center] (AWGN) {$+$};
\path[line1] (U1) -- node[at start, anchor=south,inner sep=0.1cm] {$n(t)$} (AWGN.north);
\node[block,align=center,fill=white] (WP) at (L1) {$p(t)$}; 
\node[block,align=center,fill=white] (MF) at (R2) {$p(-t)$}; 
\node[block,align=center,fill=white] (Eq) at (Equalizer) {TMVA Equalizer}; 
\path[line1] (AWGN.east) -- node[near end,anchor=south,inner sep=0.1cm] {$ $} (MF.west);
\draw [line1] (L2) -- (WP.west);
\path[line1] (WP.east) -- node[anchor=south,inner sep=0.1cm, xshift=0.15cm] {$x(t)$} (AWGN.west) ;

\node[blue, anchor=north,align=center,inner sep=0.1cm,yshift=2.15cm, xshift=-2.1cm] at (L1) {Transmitter};
\node[blue, anchor=north,align=center,inner sep=0.1cm,yshift=2.15cm, xshift=-2.45cm] at (Eq) {Receiver};

\begin{axis}[ at={(-6\hsep,-0.3\hsep)},
font=\footnotesize,
name=plot1,
axis x line=middle,
axis y line=middle,
x label style = {anchor=west},
xlabel = $t$,
ymin = -1.2, ymax = 2,
xmin = -1.5, xmax = 2.5,
ytick={},
xtick={-1,0,1,2}, xticklabels={$-T$,,$T$,$2T$},
extra x ticks={0},
extra x tick labels={$0$},
extra x tick style={xticklabel style={yshift=0.5ex, anchor=south}},
xticklabel style={xshift=0ex},
hide y axis,
ytick=\empty,
height = 3.5cm, width = 4.5cm]

\draw[->,color=black] (-1,0) -- node[at end, anchor=south]{$A_{-1}$} (-1,1);
\draw[->,color=black] (0,0) -- node[at end, anchor=north]{$A_0$} (0,-0.5);
\draw[->,color=black] (1,0) -- node[at end, anchor=south]{$A_1$} (1,0.75);
\draw[->,color=black] (2,0) -- node[at end, anchor=south]{$A_2$} (2,1.25);
\end{axis}

\draw[thick] (MF.east) -- node[anchor=south,inner sep=0.1cm] {$y(t)\, $} (Switch_L);
\path [thick] (Switch_R) -- (Switch_U);
\draw [thick] (Switch_U) -- (Switch_R);
\path[line1] (Switch_R) -- node[anchor=south,inner sep=0.1cm] {$ \tilde{A}_k \quad $} (Eq.west);

\draw [line1] (Eq.east) -- (Rend);
\draw[thick,dashed,->] (dash_U) -- node [at start, anchor=south, inner sep=0.1\hsep] {sample at $t=kT$} (dash_B);

\end{tikzpicture}
	\caption{System model under consideration. The transmit filter has a duration $T_s$ (see (\ref{eq:DefDuration})) and a bandwidth $W$ (see (\ref{eq:OOBEDef})), and the receiver is based on the TMVA from \cite{liveris2003exploiting}.}
	\label{FigSystemModel}
\end{figure*}

We consider the baseband pulse amplitude modulation system illustrated in Fig. \ref{FigSystemModel}. The transmitted symbols $\left\{A_k\right\}_{k\in\mathbb{Z}}$ are independent and identically distributed and are chosen from a constellation of size $M$ that has a zero mean and a unit average energy. After pulse shaping, the transmitted signal is 
\begin{equation*}
    x(t)=\sum_{l\in\mathbb{Z}} A_l\, p(t-lT)
\end{equation*}
where $T$ is the modulation interval. The noise $n(t)$ is additive white Gaussian with mean zero and power spectral density (PSD) $N_0/2$. 

The transmit filter $p(t)$ is assumed to be time-limited to $T_s$ seconds, i.e., 
\begin{equation}
p(t)=0 \text{ for } t\notin [-T_s/2, T_s/2],
    \label{eq:DefDuration}
\end{equation}
and to have a unit-norm (or unit energy)
\begin{equation}
\int_{-T_{s}/2}^{T_{s}/2} p^2(t)dt = 1.
\label{eq:UnitNorm}
\end{equation}
 Being time-limited, $p(t)$ cannot be strictly band-limited. We define the bandwidth of $p(t)$ by the value of $W$ that satisfies
\begin{equation}
    \int_{-W}^{W} \left| P(f)  \right|^2df=1-\epsilon,
    \label{eq:OOBEDef}
\end{equation}
where $\epsilon$ is the OOBE, and $P(f)=\int_{-T_s/2}^{T_s/2} p(t) e^{-j2\pi f t} dt$ is the Fourier transform of $p(t)$.

At the receiver, the output of the matched filter $p(-t)$ is 
\begin{equation*}
    y(t)=\sum_{l\in\mathbb{Z}} A_l \, h(t-lT) + n(t)*p(-t)
\end{equation*}
where $*$ is the convolution operation and 
\begin{equation}
 h(t)= p(t)*p(-t) = \int_{-T_s/2}^{T_s/2} p(\tau)p(\tau-t)d\tau.
 \label{eq:hConvpp}
\end{equation}
Since $p(t)$ is unit norm, then $h(0)=1$. Sampling the signal $y(t)$ at $t=kT$ yields
\begin{equation}
 \tilde{A}_k = A_k + \underbrace{\sum_{l\neq 0} A_{k-l} \, h(lT)}_{\text{ISI}} + \underbrace{n(t)*p(-t)|_{t=kT}}_{\text{colored noise}},
 \label{eq:ApVsA}
\end{equation}
 where the noise term is a zero-mean Gaussian process with variance $N_0/2$, and PSD 
\begin{equation}
 \frac{N_0}{2T}\sum_{m\in \mathbb{Z}}{ \left| P\left( \frac{f-m}{T} \right)  \right|^2} .
 \label{eq:PSDNoiseDT}
\end{equation}

In this paper, we consider pulses and modulation intervals that are capable of increasing the data rates by accepting ISI (see \eqref{eq:ApVsA}). Note that if the pulse $p(t)$ does not satisfy the Nyquist criterion, then the PSD in (\ref{eq:PSDNoiseDT}) is not flat, and hence the noise at the input of the equalizer is colored. The role of the equalizer in Fig. \ref{FigSystemModel} is to detect the transmitted sequence of  symbols in the presence of ISI and the colored noise.

Here we consider the equalizer to be the TMVA proposed in \cite{liveris2003exploiting}. In this case, the equalizer considers $2L$ interfering symbols and treats the RISI as noise. The complexity of the equalizer is proportional to $M^L$ and its performance depends on the variance of the RISI \cite{liveris2003exploiting}. The RISI is given by 
\begin{equation*}
    \text{RISI}=\sum_{|l|>L} A_{k-l}  \, h(lT),
\end{equation*}
which has zero mean and variance
\begin{equation}
    \sigma^2_{\text{RISI}} = \sum_{|l|>L} h^2(lT).
    \label{eq:SigmaRISI}
\end{equation}
In view of \eqref{eq:hConvpp} and \eqref{eq:SigmaRISI}, the complexity and performance of the equalizer depend on the used pulse $p(t)$. For a given $L$ (i.e., for a fixed complexity), the performance is maximized when using a pulse that minimizes $\sigma^2_{\text{RISI}}$. 

\section{Problem Formulation and Proposed Solution}
\label{Sec:ProbForm}

We consider the problem of finding the optimal pulse $p(t)$ that minimizes $\sigma^2_{\text{RISI}}$  in (\ref{eq:SigmaRISI}). We formulate the optimization problem as follows
\begin{equation}
    p^*(t) =   \underset{p(t)}{\text{argmin}}  \sum_{|l|>L} \left( p(t)*p(-t) |_{t=lT} \right)^2,
   \label{eq:OptProb}
\end{equation}
where $p(t)$ should satisfy (\ref{eq:DefDuration}), (\ref{eq:UnitNorm}), and (\ref{eq:OOBEDef}), and where minimizing the objective function of (\ref{eq:OptProb}) minimizes (\ref{eq:SigmaRISI}). The constraints are used to restrict the search to unit norm time-limited pulses that have an OOBE equal to $\epsilon$. In the following, we propose a numerical method to find the optimal solution of (\ref{eq:OptProb}).

Since $p(t)$ has unit energy and is $T_s$-seconds time-limited, it can be  written as a linear combination of normalized truncated PSWFs\footnote{Appendix~\ref{SecPSWFs} includes the definition and some properties of the PSWFs.}
\begin{equation}\label{aaaa}
    p(t)=\sum\limits_{i=0}^\infty \alpha_i \frac{D\varphi_{c,i}(t)}{\sqrt{\lambda_{c,i}}},
\end{equation} 
 where $c=2 T_s W$ is the time-bandwidth product of the PSWFs and $\lambda_{c,i}$ is the eigenvalue of the $i^{th}$ PSWF. Note that the PSWFs and their eigenvalues do not have closed form expressions. In this work we generate them numerically using the software in \cite{adelman2014software}.

 In view of \eqref{aaaa}, solving \eqref{eq:OptProb} is equivalent to optimizing the coefficients $\{\alpha_{i}\}_{i=0}^{\infty}$.  Optimizing numerically over the infinite set $\{\alpha_{i}\}_{i=0}^{\infty}$ is not feasible, so we approximate the optimal solution using a finite subset of the truncated PSWFs and we optimize over the set $\{\alpha_{i}\}_{i=0}^{N-1}$. We define the approximation error to be the energy of $\sum\limits_{i=N}^\infty \alpha_i \frac{D\varphi_{c,i}(t)}{\sqrt{\lambda_{c,i}}}$, and by Property \ref{property:ApproxError} in Appendix \ref{SecPSWFs}, the approximation error can be made smaller than $\epsilon$ using finite values of $N$. The resulting optimization problem at hand is
\begin{equation}
   \begin{aligned}
    \{\alpha^*_{i}\}_{i=0}^{N-1} &=  \,\, \underset{\{\alpha_{i}\}_{i=0}^{N-1}}{\text{argmin}} && \sum_{|l|>L} \left(\sum\limits_{i=0}^{N-1} \alpha_i \frac{D\varphi_{c,i}(t)}{\sqrt{\lambda_{c,i}}}* \right.\\
    & && \quad\left.\left.\sum\limits_{i=0}^{N-1} \alpha_i \frac{D\varphi_{c,i}(-t)}{\sqrt{\lambda_{c,i}}} \right|_{t=lT} \right)^2\\
   & \text{subject to}  && \sum_{i=0}^{N-1}\alpha_i^2=1\\
   &  && \sum\limits_{i=0}^{N-1}\alpha_i^2\lambda_{c,i} = 1- \epsilon
   \end{aligned},
   \label{eq:OptProbPSWFs}
\end{equation}
 where (\ref{eq:DefDuration}) is satisfied by writing $p(t)$ as a combination of truncated PSWFs, and the equivalent constraints of (\ref{eq:UnitNorm}) and (\ref{eq:OOBEDef}) are obtained using (\ref{eq:EnergyCombPSWFs}) and (\ref{eq:EnergyConcCombPSWFs}) respectively.
 
 \section{Numerical  Results}
\label{Sec:NumOpt}

\begin{figure}[t!]
\centering
\input{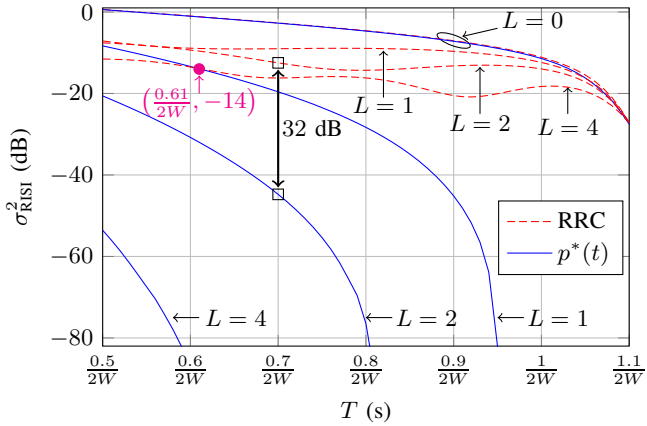}
\caption{Performance of obtained $p^*(t)$ and truncated RRC with $\beta=0.1$, for $T_s=\frac{15}{2W}$ and $\epsilon=4.4\times 10^{-4}$.}
    \label{Fig:RISI1}
\end{figure}

We chose the RRC pulses as a baseline for our system model. RRC pulses are characterized by their roll-off factor $\beta$, where the modulation interval $T=\frac{1+\beta}{2W}$ avoids the ISI in (\ref{eq:ApVsA}). As RRC pulses have infinite time duration, they are truncated to some finite time window $T_s$, which results in non-zero ISI and some OOBE. We selected $\beta=0.1$ 
and we considered a truncation window $T_s=\frac{15}{2W}$, where the obtained OOBE is $\epsilon \approx4.4 \times 10^{-4}$. In Fig. \ref{Fig:RISI1} we show the obtained $\sigma^2_{\text{RISI}}$ (which is defined in \eqref{eq:SigmaRISI}), where $T=\frac{1.1}{2W}$ is the Nyquist signaling rate, and we refer to $T<\frac{1.1}{2W}$ as the FTN signaling rates. For example, using $T=\frac{0.7}{2W}$ provides $\left(\frac{1.1}{0.7}-1\right)\times100\approx57\%$ faster transmissions which increases the data rates by $57\%$. Note that $L=0$ implies that no equalizer is used at the receiver, and as $L$ increases, the power of the RISI decreases since the equalizer considers $2L$ interfering symbols.

To have a fair comparison with the truncated RRC pulses, we use the parameters $T_s=\frac{15}{2W}$ and $\epsilon=4.4\times 10^{-4}$ in our optimization problem (\ref{eq:OptProbPSWFs}), where the time-bandwidth product of the PSWFs is $c=2T_sW =15$.
We use the MATLAB optimization function \textit{fmincon} to search for the optimal solution of (\ref{eq:OptProbPSWFs}). We performed the optimization using  different values of $N$ (the number of used truncated PSWFs). We found that the solutions converge for $N=22$ where using $N=30$ did not improve the achieved $\sigma^2_{\text{RISI}}$, which is consistent with the analysis in Sec. \ref{Sec:ProbForm} where we showed that the optimal solution can be approximated using a finite value of $N$. In Fig. \ref{Fig:RISI1} we present the obtained $\sigma^2_{\text{RISI}}$ of the optimal pulses  for $\epsilon=4.4 \times 10^{-4}$, $T\in\left[\frac{0.5}{2W}, \frac{1+\beta}{2W}\right]$ with a step of $\frac{0.01}{2W}$, and $L\in\{0,1,2,4\}$. For $L=0$, the truncated RRC pulse is very close to the optimal one. For $L>0$, the obtained pulses show a large decrease in the RISI variance. This allows for example to use the optimal pulses with $L=1$ instead of the truncated RRC with $L=4$ for $T\geq 0.61/(2W)$ and $\sigma^2_{\text{RISI}}\leq -14$~dB. 
We also considered comparing with the RRC pulse with $\beta=0.2$, where the obtained OOBE is $\approx 9.5\times10^{-5}$ for $T_s=\frac{15}{2W}$. We do not include the obtained results here since they have the same pattern as the ones in Fig. \ref{Fig:RISI1}.

\begin{figure*}[t!]
\centering
\input{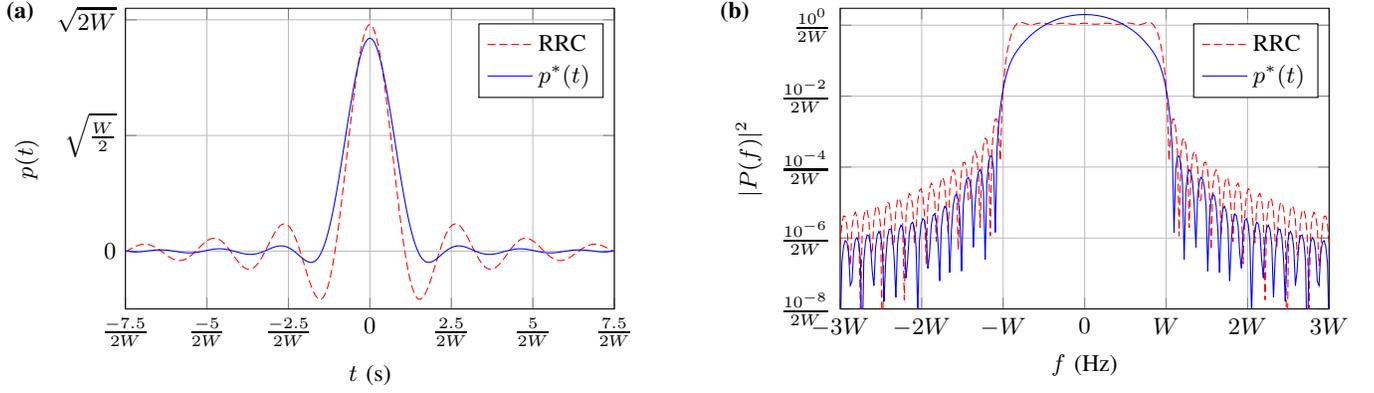}
\caption{(a) Optimal pulse versus RRC at `{\tiny$\square$}' in Fig. \ref{Fig:RISI1}, and  (b) their corresponding $|P(f)|^2$.}
    \label{Fig:pulse}
\end{figure*}

\begin{figure*}[t!]
\centering
\input{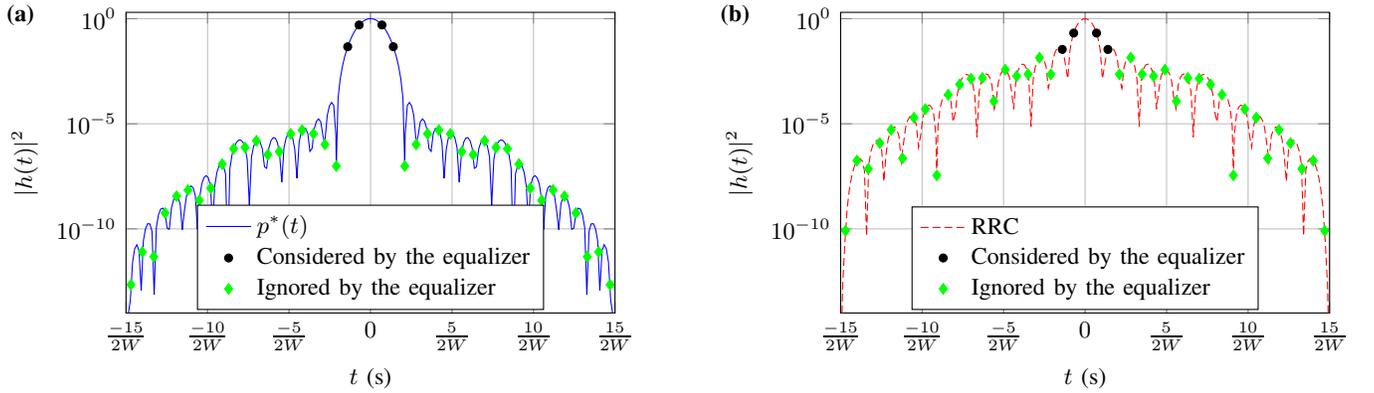}
\caption{(a) $|h(t)|^2$ corresponding to $p^*(t)$ in Fig. \ref{Fig:pulse}, and (b) $|h(t)|^2$ corresponding to RRC of Fig. \ref{Fig:pulse}. The equalizer considers $2L=4$ interfering symbols.}
    \label{Fig:Hpulse}
\end{figure*}

In Fig. \ref{Fig:pulse}a we show the obtained optimal pulse and the RRC pulse at `{\tiny$\square$}' in Fig. \ref{Fig:RISI1}, i.e. for $L=2$ and  $T=0.7/(2W)$. We also show their frequency spectrum in Fig. \ref{Fig:RISI1}b. In Fig. \ref{Fig:Hpulse} we show the corresponding $|h(t)|^2$ of the pulses of Fig. \ref{Fig:pulse}, where $h(t)$ is defined in (\ref{eq:hConvpp}). In table \ref{TableCoef} we show the coefficients of the PSWFs for the optimal pulse $p^*(t)$ of Fig. \ref{Fig:pulse}, where the presented $\{\alpha_i^*\}_{i=0}^{N-1}$ are the solution of (\ref{eq:OptProbPSWFs}) for the design parameters $\epsilon=4.4\times 10^{-4}$, $T=0.7/(2W)$, and $L=2$.  We found that the optimal pulse changes (i.e. $\{\alpha_i^*\}_{i=0}^{N-1}$ change) when changing the value of one of the design parameters ($\epsilon$, $T$, and $L$). We also noticed that the obtained optimal pulses are symmetric, which means that we can write them as a linear combination of only symmetric truncated PSWFs, i.e., the ones with even index $i$ (see Appendix \ref{SecPSWFs}). Therefore we can reduce the complexity of numerical optimization by optimizing (\ref{eq:OptProbPSWFs}) over the $\alpha_i$ with even index $i$, while fixing $\alpha_i=0$ for odd indices.

\begin{table*}[t!]
    \centering
    \caption{The coefficients $\{\alpha_i^*\}_{i=0}^{21}$  for the optimal pulse $p^*(t)$ of Fig. \ref{Fig:pulse}.}
    \begin{tabular}{|c|c|c|c|c|c|c|c|c|c|c|c|}
    \hline  $\alpha^*_0$  & $\alpha^*_2$  & $\alpha^*_4$  & $\alpha^*_6$  & $\alpha^*_8$  & $\alpha^*_{10}$  & $\alpha^*_{12}$  & $\alpha^*_{14}$  & $\alpha^*_{16}$  & $\alpha^*_{18}$  & $\alpha^*_{20}$  & $\alpha^*_i$ for odd $i$  \\
    \hline    $0.8053$&   $-0.442$&  $0.2923$&  $-0.1996$&  $0.136$&  $-0.0905$&  $0.0562$&  $-0.03$&  $0.0107$&  $0.00014$&  $0.0011$ & $0$ \\
    \hline
    \end{tabular}
    \label{TableCoef}
\end{table*}

\begin{figure}[t!]
\centering
\begin{tikzpicture}[font=\small,
    Gclabel/.style={text=black},]
\begin{semilogyaxis}[
    xlabel={$E_b/N_0$ (dB)},
    ylabel={BER},
    xmin=6, xmax=38,
    ymin=3.25e-6, ymax=0.1,
    legend style={at={(0.68,0.04)},anchor=south},
     xtick={6,10,14,18,22,26,30,34,38},
    ytick={1e-5, 1e-4,1e-3,1e-2,1e-1},
    ymajorgrids=true,
    xmajorgrids =true,
    yminorgrids=false,
    xminorgrids =false,
    legend cell align={left},    
    scale only axis,
    width=7cm,
    height=5cm,
]

\draw[black!75, rotate around={-17:(30,5.5e-4)}](30,5.5e-4) ellipse  (8 and 1.75);
\draw[black!75, rotate around={-13.5:(28,5e-2)}](28,5e-2) ellipse  (9.5 and 1.75);
\draw[black, ->] (29,5e-3) -- (30,1.1e-3) node[pos=-0.2] {Error floor caused by RISI};
\draw[black, ->] (29,0.9e-2) -- (28,3e-2) node[pos=-0.2] {$ $};

\addplot[color=red,
    dashed,
    dash pattern=on 3pt off 1.5pt,
    mark=square,mark size=1.5pt,
    mark options={solid},
    ]
coordinates{(6,9.065850e-02)(8,7.455305e-02)(10,6.486805e-02)(12,6.024535e-02)(14,5.481882e-02)(16,5.121597e-02)(18,5.188400e-02)(20,4.961376e-02)(22,4.903262e-02)(24,4.903106e-02)(26,4.952406e-02)(28,4.860110e-02)(30,4.831287e-02)(32,4.983766e-02)(34,4.898234e-02)(36,4.799074e-02)(38,4.938919e-02)(40,4.819816e-02)  };\addlegendentry{RRC, $4$-state TMVA}

\addplot[
    color=red,
    dashed,
    dash pattern=on 3pt off 1.5pt,
    mark=o,mark size=1.5pt,
    mark options={solid},
    ]
    coordinates {(6,4.266436e-02)(8,2.071074e-02)(10,9.427611e-03)(12,4.613356e-03)(14,2.516352e-03)(16,1.494328e-03)(18,1.111592e-03)(20,7.840039e-04)(22,6.375175e-04)(24,5.968394e-04)(26,5.281411e-04)(28,5.009701e-04)(30,4.669712e-04)(32,4.552923e-04)(34,4.627043e-04)(36,4.471797e-04)(38,4.837524e-04)(40,4.493809e-04)
    };
\addlegendentry{RRC, $16$-state TMVA}

\addplot[
    color=red,
    dashed,
    dash pattern=on 3pt off 1.5pt,
    ]
    coordinates {
(6,3.007058e-02)(8,9.746622e-03)(10,2.426020e-03)(12,5.203760e-04)(14,1.015986e-04)(16,2.133651e-05)(18,3.070711e-06)(20,0)(22,0)(24,0)(26,0)(28,0)(30,0)(32,0)(34,0)(36,0)(38,0)(40,0) 
    };
\addlegendentry{RRC, $128$-state TMVA}

\addplot[
    color=blue,
    ]
    coordinates {(6,4.135974e-02)(8,1.084489e-02)(10,1.306370e-03)(12,8.117124e-05)(14,3.693636e-06)
    };
\addlegendentry{$p^*(t)$, $4$-state TMVA}

\end{semilogyaxis}    
\end{tikzpicture}



\caption{BER simulations of the pulses of Fig. \ref{Fig:pulse}(a).}
    \label{Fig:BER}
\end{figure}
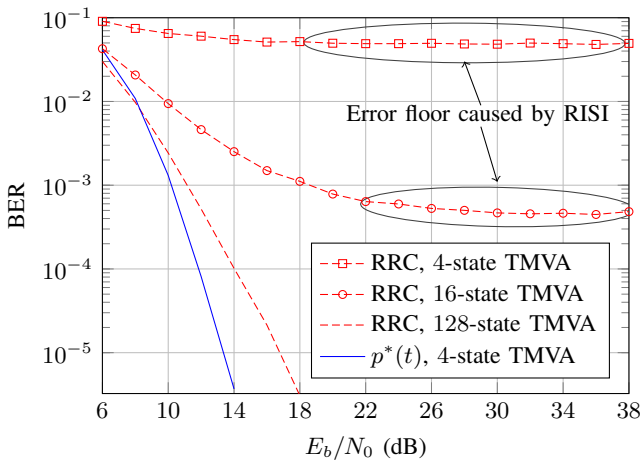

To conclude, we performed numerical simulations to evaluate the bit error rate (BER) of our system. We considered a binary input constellation (i.e., $A_k\in\{-1,1\}$), the modulation interval $0.7/(2W)$,  and the pulses of Fig. \ref{Fig:pulse}. For the optimal pulse we used  a TMVA equalizer with $2^L=4$ states, and we used $4$-state, $16$-state, and $128$-state TMVA equalizers for the truncated RRC pulse (i.e., using $L=2$, $L=4$, and $L=7$). Fig.~\ref{Fig:BER} shows that the obtained optimal pulse with $4$-state TMVA outperforms the truncated RRC with $128$-state TMVA in terms of BER, where $E_b$ is the average energy per bit ($E_b=1$ in our case) and $N_0$ determines the PSD of the noise. Hence, the optimal pulse can simultaneously improve the BER performance and reduce the complexity of the receiver. Note that the truncated RRC suffers from an error floor where the RISI limits the BER for very low noise power (or equivalently high $E_b/N_0$).

\section{Conclusions}
\label{Sec:Conc}

The main conclusion of this paper is that optimizing the pulse shaping filters is essential for the practical implementations of FTN systems, where we showed that the optimal pulses can simultaneously improve the performance and decrease the complexity of truncated equalizers. We also found that there is no universal optimal filter, where we obtained different optimal pulses for the different design parameters: the modulation interval, the out-of-band energy, and the number of interfering terms considered by the equalizer.

\appendix

\subsection{The Prolate Spheroidal Wave Functions}
\label{SecPSWFs}

For any positive time-bandwidth product $c=2 T_s W$, the PSWFs $\{\varphi_{c,i}(t) \}_{i=0}^{\infty}$ are the normalized solutions of the integral equation \cite{slepian1961prolate}\footnote{Note that our definition of the time-bandwidth product is different than the one in \cite{slepian1961prolate}. For example, $\varphi_{1,i}(t)$ here is the same as $\varphi_{\frac{\pi}{2},i}(t)$ in \cite{slepian1961prolate}.}
\begin{equation*}
  \lambda_{c,i} \, \varphi_{c,i}(t) = \int_{-T_s/2}^{T_s/2} \frac{\sin{2\pi W(t-s)}}{\pi(t-s)}
  \varphi_{c,i}(s) \, ds,
\end{equation*}
where $\lambda_{c,i}$ is the eigenvalue of $\varphi_{c,i}(t)$. The PSWFs are real band-limited continuous functions that are symmetric when $i$ is even and antisymmetric when $i$ is odd.

We denote by $D \varphi_{c,i}(t)$ the truncated PSWF in the time window $[-T_s/2, T_s/2]$. The energy of $D \varphi_{c,i}(t)$ is equal to $\lambda_{c,i}$ \cite{slepian1961prolate}. Hence the normalized truncated PSWFs are
\begin{equation*}
    \left\{\frac{D\varphi_{c,i}(t)}{\sqrt{\lambda_{c,i}}}\right\}_{i=0}^{\infty},
\end{equation*}
and they form a complete orthonormal set for $T_s$-seconds time-limited functions that have finite energies.

Consider a unit norm $T_s$-seconds time-limited signal $p(t)$, then $p(t)$ can be uniquely written as a linear combination of the normalized truncated PSWFs as given in (\ref{aaaa}), i.e. 
$p(t)=\sum_{i=0}^{\infty}\alpha_i\frac{D\varphi_{c,i}(t)}{\sqrt{\lambda_{c,i}}}$.
By Parseval, the energy of $p(t)$ depends only on the coefficients $\left\{\alpha_{i}\right\}_{i=0}^{\infty}$ and it is equal to
\begin{equation}
    \int_{-T_s/2}^{T_s/2}|p(t)|^2dt=\sum_{i=0}^{\infty}\alpha_i^2.
    \label{eq:EnergyCombPSWFs}
\end{equation}
The eigenvalue $\lambda_{c,i}$ is equal to the energy of $\frac{D\varphi_{c,i}(t)}{\sqrt{\lambda_{c,i}}}$ in the bandwidth $[-W, \, W]$, and in the following property we derive the energy of $p(t)$ in the bandwidth $[-W, \, W]$.

\begin{property}
\label{property:OOBE}
The energy of $p(t)$ in the frequency band $[-W, \, W]$ is equal to
\begin{equation}
    \int_{-W}^{W}|P(f)|^2 df=\sum_{i=0}^{\infty}\alpha_i^2\lambda_{c,i}
    \label{eq:EnergyConcCombPSWFs}
\end{equation}
where $P(f)$ is the Fourier transform of $p(t)$.
\end{property}
\begin{proof}
The Fourier transform of a truncated PSWF is equal to a scaled version of the PSWF 
\cite[eq. (4)]{jaffal2020time}. Therefore the Fourier transforms of the truncated PSWFs are orthogonal over $[-W, \, W]$, and hence the left-hand side of (\ref{eq:EnergyConcCombPSWFs}) is equal to the sum of the energies of $\alpha_i\frac{D\varphi_{c,i}(t)}{\sqrt{\lambda_{c,i}}}$ in the band $[-W, \, W]$.
\end{proof}

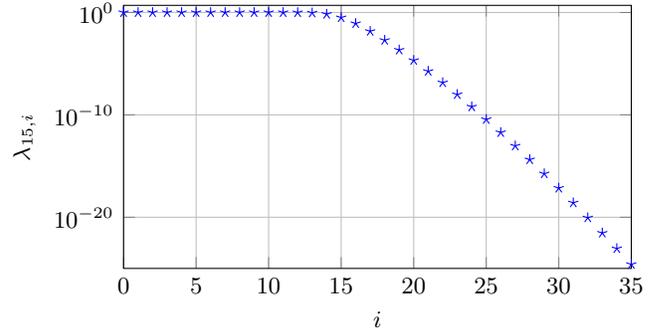
\begin{figure}[t!]
\centering
\begin{tikzpicture}[font=\small]
\begin{semilogyaxis}[
    xlabel={$i$},
    ylabel={$\lambda_{15,i}$},
    xmin=0, xmax=35,
    ymin=1e-25, ymax=5,
    xtick={0,5,10,15,20,25,30,35},
    ytick={1e-20, 1e-10, 1},
    ymajorgrids=true,
    xmajorgrids =true,
    yminorgrids=true,
    xminorgrids =true,
    scale only axis,
    width=6.75cm,
    height=3.5cm,
]

\addplot[blue, mark =star,  only marks]
    coordinates {
(0 , 1.000000e+00) (1 , 1.000000e+00) (2 , 1.000000e+00) (3 , 1.000000e+00) (4 , 1.000000e+00) (5 , 1.000000e+00) (6 , 1.000000e+00) (7 , 9.999999e-01) (8 , 9.999989e-01) (9 , 9.999842e-01) (10 , 9.998155e-01) (11 , 9.982042e-01) (12 , 9.858268e-01) (13 , 9.160469e-01) (14 , 6.814855e-01) (15 , 3.175837e-01) (16 , 8.428167e-02) (17 , 1.458720e-02) (18 , 1.944676e-03) (19 , 2.175537e-04) (20 , 2.116246e-05) (21 , 1.822092e-06) (22 , 1.404612e-07) (23 , 9.777509e-09) (24 , 6.187856e-10) (25 , 3.580493e-11) (26 , 1.903371e-12) (27 , 9.334581e-14) (28 , 4.238922e-15) (29 , 1.788262e-16) (30 , 7.029222e-18) (31 , 2.581372e-19) (32 , 8.878272e-21) (33 , 2.866306e-22) (34 , 8.704425e-24) (35 , 2.491285e-25) (36 , 6.732210e-27) (37 , 1.720586e-28) 

    };

\end{semilogyaxis}    
\end{tikzpicture}
\caption{Eigenvalues of the PSWFs for $c=15$ and $i=0,1,\dots, 35$}
    \label{Fig:Lambdas10}
\end{figure}

The eigenvalue $\lambda_{c,i}$ decreases as $i$ increases and  $1>\lambda_{c,i}>\lambda_{c,i+1}>0$ for all $i\geq 0$. For relatively low index $i$ (i.e., $i<<c$), the eigenvalues are very close to $1$. And as $i$ increases beyond $c$, $\lambda_{c,i}$ decreases exponentially towards zero. As an example, Fig. \ref{Fig:Lambdas10} shows the eigenvalues for $c=15$ where a logarithmic scale is used on the y-axis. The exponential decay of $\lambda_{c,i}$ allows to approximate the sum in (\ref{eq:EnergyConcCombPSWFs}) using a finite number of terms, and as shown in the following property, it also allows to approximate $p(t)$ using a finite number of truncated PSWFs if the OOBE of $p(t)$ is very small.

\begin{property}
\label{property:ApproxError}
If $\int_{-W}^{W}|P(f)|^2 df=1-\epsilon$,
then there exists a finite integer $N$ where an upper bound on the energy of $\sum_{i=N}^{\infty}\alpha_i\frac{D\varphi_{c,i}(t)}{\sqrt{\lambda_{c,i}}}$ is well approximated by $\epsilon$.
\end{property}
\begin{proof}
Using (\ref{eq:EnergyCombPSWFs}) and (\ref{eq:EnergyConcCombPSWFs}), the OOBE is given by
\begin{equation*}
    \sum\limits_{i=0}^{\infty}\alpha_i^2(1-\lambda_{c,i}) = \epsilon .
\end{equation*}
For any integer $N>c$, $(1-\lambda_{c,i}) > (1-\lambda_{c,N-1})>0$ $\forall i \geq N$. For sufficiently large $N$, $(1-\lambda_{c,N-1})\approx 1$ since $\lambda_{c,N-1}$ decays exponentially with $N$ (see the example in Fig. \ref{Fig:Lambdas10}). Then 
\begin{align*}
      \epsilon  &=\sum\limits_{i=0}^{N-1}\alpha_i^2(1-\lambda_{c,i})+\sum\limits_{i=N}^\infty \alpha_i^2(1-\lambda_{c,i})\\
     \epsilon  &> \sum\limits_{i=N}^\infty \alpha_i^2(1-\lambda_{c,N-1})\\
    \epsilon   &\approx \epsilon /(1-\lambda_{c,N-1}) > \sum\limits_{i=N}^\infty \alpha_i^2,
\end{align*}
where $\sum\limits_{i=N}^\infty \alpha_i^2$ is the energy of $\sum_{i=N}^{\infty}\alpha_i\frac{D\varphi_{c,i}(t)}{\sqrt{\lambda_{c,i}}}$.
\end{proof}

\ifCLASSOPTIONcaptionsoff
  \newpage
\fi

\bibliographystyle{ieeetr}
\bibliography{references}

\end{document}